\documentstyle{article}

\def\beg{\begin{equation}} \def\ene{\end{equation}}

\begin{document}

\large

$$   $$ 
\vskip 5cm \Large 
\centerline{\bf Heaviside transform
of the effective
potential} \vskip 14pt 
\centerline{\bf  in the Gross-Neveu
model} \vskip
1.5cm
\large \centerline{\sc Hirofumi Yamada} \vskip 1cm
\centerline{\it
Department of
Mathematics, Chiba Institute of Technology} \vskip 10pt
\centerline{\it
2-1-1
Shibazono, Narashino-shi, Chiba 275} \vskip 10pt
\centerline{\it Japan}
\vskip
10pt \centerline{\it e-mail:yamadah@cc.it-chiba.ac.jp} \vskip
2cm
\baselineskip
18pt \large \centerline{\bf Abstract} \vskip 10pt
\large
Unconventional way of handling the perturbative series is presented 
with the help of Heaviside transformation with respect to the
mass.  We apply Heaviside transform to the
effective potential in the massive Gross-Neveu model and
carry out perturbative
approximation of the massless potential by dealing with the
resulting Heaviside
function.  We find that accurate values of the dynamical mass
can be obtained
from the Heaviside function already at finite orders where
just the several 
of diagrams are incorporated.  We prove that our approximants
converges
to the exact massless potential in the infinite order.  
Small mass expansion of the effective potential can be also
obtained in our approach.
\newpage \baselineskip 24pt

\noindent {\bf  1 Introduction}

\vskip 10pt 
Even if the proof of dynamical massless symmetry breaking
requires
genuine non-perturbative approaches, it does not necessarily
mean that
the
perturbative expansion is totally useless.  There is the
possibility
that
non-perturbative quantities in the massless limit may be
approximately
calculated
via perturbative approach.  The purpose of this paper is to
explore the
possibility and show a concrete affirmative result by
re-visiting the
Gross-Neveu
model${^{1}}$.  

Let us consider the effective potential of the Gross-Neveu
model.  
As is well known, ordinary massless perturbation expansion
gives
infra-red
divergences and to cure the problem one must sum up all the
one-loop
diagrams.  Then the summed result reveals the non-trivial
vacuum
configuration of $<\bar \psi \psi>$ and the dynamical
generation of the
mass.  

The point we like to note is whether such a non-perturbative
effect
needs, in
the approximate evaluation, the infinite
sum of perturbative contributions.  To
resolve the issue, we deal with a truncated series
$V_{pert}$, without
conventional loop
summation, and study the approximate calculation of the
effective
potential $V$ 
at $m=0$.  

A naive way of approximation would go as the following: To
get around
the
infrared singularity we turn to the massive
case and probe $V_{pert}(\sigma,m)$ at small $m$.  Since the
limit,
$m\rightarrow 0$, cannot be taken
in $V_{pert}(\sigma,m)$, we may choose some non-zero $m$
($=m^{*}$) and approximate the effective potential $V(\sigma,
m=0)$ by
$V_{pert}(\sigma, m^{*})$.  However, the problem is that 
$V_{pert}(\sigma, m)$
is not valid for small enough $m$.  
This is the place where the Heaviside function comes in.  Our
suggestion to
resolve the problem is to contact the 
Heaviside transformation of $V(\sigma,m)$ with respect to the
mass$^{2,3}$.  

Heaviside transform of the effective potential, $\hat V$, is
a function
of $\sigma$ and $x$ which is conjugate with $m$.  Then, the
key relation
is that
$\lim_{m\rightarrow 0}V(\sigma, m)=\lim_{x\rightarrow
\infty}\hat
V(\sigma, x)$.  Of course this is valid only when the both
limits exist and do
not apply for $V_{pert}$ and its Heaviside function, $\hat
V_{pert}$, because
those functions diverge in the limits.   However there arises
the possibility
that $\hat V(\sigma,\infty)$ and hence $V(\sigma,0)$ may be
well approximated by
putting some finite value of $x$ into $\hat V_{pert}$.  This
is because $\hat V_{pert}$ has the convergence radius much
larger than that of $V_{pert}$.  
Although $\hat V_{pert}$ shares the similar infra-red 
problems with $V_{pert}$, we will find that $\hat V_{pert}$
is much more 
convenient in this kind of massless approximation.  
 Actually we will demonstrate that, at finite
perturbative orders where just the several of Feynman
diagrams are
taken into account, the accurate dynamical mass
is obtained
via the Heaviside transform approach. 

Throughout this paper, we use dimensional
regularization$^{4}$.  We
confine
ourselves with the leading order of large $N$ expansion and
$N$ is
omitted
for
the sake of simplicity. 
\vskip 20pt 
\noindent {\bf 2. Heaviside transform with
respect to the mass}

In this section we summarize basic features of the Heaviside
transform
and illustrate our strategy by taking a simple example. 

 Let
$\Omega(m)$ be a given function of the mass $m$.  The
Heaviside
transform of $\Omega(m)$ is given by the Bromwich integral, 
\begin{equation} 
\hat \Omega(x)=\int^{s+i\infty}_{s-i\infty}{dm \over 2\pi
i}{\exp(m x)
\over
m}\Omega(m), 
\end{equation} 
where the vertical straight contour should lie in the
right of all the possible poles and the cut of $\Omega(m)/m$
(In (1),
the real
parameter $s$ specifies the location of the contour).  Since
$\Omega(m)/m$ is
analytic in the domain, $Re(m)>s$, $\hat \Omega(x)$ is zero
when $x<0$.
It is known that the Laplace transformation (of the second
kind) gives
the original function as, 
\begin{equation}
\Omega(m)=m\int^{\infty}_{-\infty}dx\exp(-m x)\hat\Omega(x).
\end{equation} 
Since $\hat \Omega(x)=0$ for $x<0$, the
region of the integration effectively reduces to
$[0,\infty)$.  It is
easy to derive the relation, 
\begin{equation} 
\lim_{m\rightarrow
+0}\Omega(m)=\lim_{x\rightarrow +\infty} \hat\Omega(x), 
\end{equation} 
where
the both limits are assumed to exist.  As noted before, the
point of our
scheme
consists in
utilizing $\hat \Omega$ to approximate the massless value of
$\Omega$,
$\Omega(0)$, by relying upon 
(3).

To illustrate our strategy based on (3), let us
consider a simple example.  Given a following truncated
series in $1/m$,
\beg
f_{L}(m)=\sum^{L}_{n=0}{(-1)^{n} \over m^{n+1}},
\ene
we try to approximate the value of
$f(m)=f_{\infty}(m)=(1+m)^{-1}$ at
$m=0$, $f(0)=1$, by using information just contained in the
truncated
series (4).  Since the convergence radius, $\rho$, of
$f_{\infty}(m)$ is
unity, we cannot have approximation better than $1/2$ from
$f_{L}(m)$. 
However, the state changes if we deal with its Heaviside
function.  

The Heaviside transform of $f_{L}(m)$ is given by
\beg
\hat
f_{L}(x)=\sum^{L}_{n=1}(-1)^n\int^{s+i\infty}_{s-i\infty}
{dm \over 2\pi i}{\exp(m x) \over m}
{1 \over m^{n+1}}=\sum^{L}_{n=0}(-1)^n
{x^{n+1} \over (n+1)!}\theta(x),
\ene
where
\begin{equation} 
\theta(x)=\left\{ \begin{array}{@{\,}ll} 1 &
\mbox{$(x>0)$}\\ 0 & \mbox{$(x<0)$.} 
\end{array} \right.
\end{equation}
From (5) it is easy to find that $\hat
f(x)=(1-e^{-x})\theta(x)$ and (3)
holds for $f$ and $\hat f$.  For our purpose it is crucial
that
$\rho=\infty$ for $\hat f_{\infty}$ while $\rho=1$ for
$f_{\infty}$.  The infinite convergence radius ensures us to
probe the
large $x$ behavior of $\hat f$ by $\hat f_{L}$ to arbitrary
precision by increasing perturbative order.    
Due to the truncation, however, $\hat f_{L}$ diverges as
$x\rightarrow
\infty$.  Then, in approximating
$\hat f(\infty)$ and therefore $f(0)$, we stop taking the
limit and input some
finite value into $x$.  The input value of $x$, say $x^{*}$,
should be
taken as large as
possible in the reliable perturbative region in $x$.  At this
place we
understand that the good convergence property of $\hat f_{L}$
is one of
the advantage of Heaviside function.  

Since the upper limit
of
perturbative region is not a rigorously defined concept, we
determine
the input value $x^{*}$ in heuristic way.  Our suggestion to
fix $x^{*}$ is as the 
following:  The series (5) is valid for small $x$ but breaks
down at large $x$.  The breaking appears as the
domination of the highest term in $\hat f_{L}$ which leads to
the unlimited
growth or decreasing of the function (see Fig.1).  Thus
$x^{*}$ is located
somewhere around the beginning of the dominating behavior. 
Then, for odd
$L$ and large even $L$, we find the plateau region just before
the domination and
that the region represents the end of the
perturbative regime.  Thus, we choose the stationary point in
the plateau
region as representing the typical violation of
the perturbation expansion.   Hence we fix
$x^{*}$ by the stationarity condition,
\beg
{\partial \hat f_{L}(x^{*}) \over \partial x^{*}}=0.
\ene
The condition (7) reads as
\beg
\theta(x^{*})\sum^{L}_{n=0}{(-x^{*})^n \over
n!}+\delta(x)\sum^{L}_{n=0}{(-1)^n (x^{*})^{n+1} \over
(n+1)!}=0,
\ene
and reduces to
\beg
\sum^{L}_{n=0}{(-x^{*})^n \over n!}=0.
\ene
The solution exists for odd $L$ and it varies with $L$.  We
find from
(9)
that the solution $x^{*}$ tends to $\infty$ as $L\rightarrow
\infty$. 
More
precisely the solution scales for large $L$ as
\beg
x^{*}\sim {1 \over 3}L.
\ene

We have explicitly done the numerical experiment to several
higher
orders and 
obtained the result for $L=1,5,9,13,17$,
\beg
\hat f_{L}(x^{*})=0.5,\hskip 4pt 0.850675,\hskip 4pt
0.953301,
\hskip 4pt 0.985166,\hskip 4pt 0.995251,
\ene
at $x^{*}=1, 2.18061, 3.33355, 4.47541, 5.6112$,
respectively.
Thus, the sequence gives good approximation of the exact
value.  In this
toy
model, one finds that $\hat f_{L}(x^{*})$ converges in the
$L\rightarrow
\infty$
limit by using the scaling relation (10).

Up to now we have concentrated on approximating the massless
value.  We
here
point out that our scheme is
capable of constructing the small $m$ expansion of
$\Omega(m)$, that is,
the
scheme allows the approximation of the function itself when
$m$ is
small.  
Consider in general the approximation of the derivatives at
$m=0$, 
\beg
\Omega^{(k)}(0)={\partial^{k}\Omega \over \partial
m^{k}}\biggl|_{m=0}, 
\ene
which is needed when one constructs the small $m$ expansion
of
$\Omega(m)$, 
\beg 
\Omega(m)=\Omega(0)+{\Omega^{(1)}(0) \over
1!}m+{\Omega^{(2)}(0) \over 2!}m^{2}+\cdots. 
\ene 
The coefficients,
$\Omega^{(k)}(0)\hskip 3pt (k=1,2,3,\cdots)$, can be
approximated as
follows.  The starting formula is that 
\beg 
\Omega^{(k)}(m)\stackrel{{\cal
H}}{\rightarrow}\int^{x}_{-\infty}dt(-t)^{k}{\partial \hat
\Omega(t)
\over
\partial t}\def\alpha_{k}(x), 
\ene 
where we used
\beg m{\partial \Omega \over \partial m}\stackrel{{\cal
H}}{\rightarrow}
-x{\partial \hat \Omega(x) \over \partial x},\quad {1 \over
m}\Omega\stackrel{{\cal
H}}{\rightarrow}\int^{x}_{-\infty}dt\hat
\Omega(t).
\ene 
Here ${\cal H}$ above the arrow represents the Heaviside
transformation.  
Hence, from the agreement condition (3) we find
\beg
\lim_{m\rightarrow 0}\Omega^{(k)}(m)=\lim_{x\rightarrow
\infty}\alpha_{k}(x)=\int^{\infty}_{-\infty}dt(-t)^k{\partial
\hat\Omega(t)
\over \partial t}.
\ene

Now in our perturbative approach we use $\Omega_{L}$ (the
truncated
series at
the order $L$) for real $\Omega$.  Then, we show that 
we can simulate $\alpha_{k}(\infty)$ by
$\alpha_{k}(x^{*}_{k})$ where
$x^{*}_{k}$ may be fixed following the same reasoning we
presented for
the
case of $f(0)$ approximation.  Namely we guess that the break
down of
the perturbative
expansion is represented by the plateau region, if it exists,
just before
the
unlimited growth of the size of the function.  Therefore we
use stationarity condition which reads,
\beg 
{\partial \alpha_{k} \over \partial
x}\biggl|_{x=x^{*}_{k}}=(-x^{*}_{k})^{k}{\partial \hat
\Omega_{L} \over
\partial x^{*}_{k}}=0,
\ene 
and find that $x^{*}_{k}$ satisfies the same condition as
that for
$x^{*}$.  
  Thus the solution of (17) is universal for all $k$ and
fixes the
coefficients
of the small $m$ expansion
to all orders.  This is desirable since the uncertainty
connected with
the
choice
of $x^{*}$ and $x^{*}_{k}$ is minimized.  We note that for
$\alpha_{k}(x^{*})$ $(k=1,2,3,\cdots)$ the
integration is necessary and $\theta(x)$ and $\delta$
functions should
be
kept in the integrand in general.  

As an example let us calculate the small $m$
expansion of $f(m)$.  The coefficient $\alpha_{k}$ is given
at order $L$
by,
\beg
\alpha_{k}=(-1)^k\sum^{L}_{n=0}{(-1)^n x^{k+n+1} \over
n!(k+n+1)}.
\ene
By substituting $x^{*}$ at $L=17$ into $\alpha_{k}$, we have
the
following satisfactory approximant of
$f(m)=1-m+m^{2}-\cdots$;
\beg
0.995251-0.970003 m+0.90271 m^2-0.78284 m^3+0.62233
m^4+\cdots.
\ene

\vskip 20pt \noindent {\bf 3 Application to the effective
potential}
\vskip
10pt
Having prepared basic analysis, we turn to a model field
theory which is
of our
main interest.  
Consider the Gross-Neveu model at the leading order of large
$N$
expansion$^{1}$.
The Lagrangian is given within dimensional
regularization$^{4}$ at
$D=4-2\epsilon$ by \begin{eqnarray} {\cal L}&=& \bar
\psi(i\gamma^{\mu}\partial_{\mu}-m)\psi-{1 \over
2}\sigma^{2}-{g \over
\sqrt{N}}\mu^{\epsilon}\sigma \bar\psi\psi +{\cal
L}_{ct},\nonumber\\
{\cal
L}_{ct}&=&A\sigma-B{1 \over 2}\sigma^{2}, \end{eqnarray}
where
\begin{equation}
\psi=(\psi_{1},\cdots,\psi_{N}), \hskip 3mm
A=-{\sqrt{N}mg\mu^{-\epsilon}
\over
2\pi}{\hat{1 \over \epsilon}}, \hskip 3mm B={g^{2} \over
2\pi}{\hat{1
\over
\epsilon}}, \hskip 3mm {\hat{1 \over \epsilon}}={1 \over
\epsilon}-\gamma+\log(4\pi). \end{equation} Here ${\overline
{MS}}$
scheme$^{5}$
was used for the subtraction.  It is well known that the
model generates
the
dynamical fermion mass, $m_{dyn}=\Lambda$, where $\Lambda$
denotes the
renormalization group invariant scale in ${\overline {MS}}$
scheme.

At the leading order of $1/N$ expansion, the effective
potential is
given by
the
sum of diagrams shown in Fig.2.  The straightforward
calculation gives
\beg
V(\sigma, m)={m^{2} \over 4\pi}(\log{m^{2} \over
\mu^{2}}-1)+{mg\sigma
\over
2\pi}\log{m^{2} \over \mu^{2}}+{g^{2}\sigma^{2} \over
4\pi}(\log{m^{2}
\over
\Lambda^{2}}+2)-\sum^{\infty}_{n=3}{(-g\sigma)^{n} \over \pi
n(n-1)(n-2)}m^{-n+2}. \ene We note that although the naive
power
counting
with
respect to $N$ leads that the contribution with many
$\sigma$-legs
corresponds to
higher order in $1/N$, they must be included since the vacuum
value of
$\sigma$
is of order $\sqrt{N}$.

The series (22) converges only when $|g\sigma/m|<1$ and hence
the small
$m$
behavior relevant to the dynamical mass generation is not
known from
(22).
However, Heaviside transformation enlarges the convergence
radius and
enables us to
study the large $x$ behavior of the corresponding Heaviside
function,
$\hat
V(\sigma,x)$, as we can see below.

To obtain $\hat V(\sigma,x)$ we need to know the transform of
$m^{k}(k=0,1,2,\cdots), m\log m, m^{2}\log m$ and $1/m^{k}$. 
Here the
following formula is basic, 
\begin{equation}
m\Omega(m)\stackrel{{\cal H}}{\rightarrow} {\partial \hat
\Omega(1/x)
\over
\partial x}.
\end{equation} 
  For example from (1)
we have
\begin{equation} \log(m)\stackrel{{\cal
H}}{\rightarrow}(-\gamma-\log(x))\theta(x). 
\end{equation} 
The use of (23) on (24)
then leads to 
\begin{eqnarray} 
m\log m  &\stackrel{{\cal H}}{\rightarrow}& -{1
\over x}\theta(x)-(\gamma+\log  x )\delta(x),\\ m^{2}\log m 
&\stackrel{{\cal
H}}{\rightarrow}& {1 \over x^{2}}\theta(x)-{2 \over
x}\delta(x)-(\gamma+\log 
x
)\delta^{'}(x). 
\end{eqnarray} 
The transformation of $m^{k}$ is easily obtained
from 
\begin{equation} 
1 \stackrel{{\cal H}}{\rightarrow} \theta(x),
\end{equation} 
as 
\begin{equation} m^{k} \stackrel{{\cal H}}{\rightarrow}
\delta^{(k-1)}(x)\quad (k=1,2,3,\cdots). 
\end{equation} The $\delta$ functions
are needed when one carries out Laplace integrals for the
Heaviside
functions.
This is because the $\delta$ function terms cancel out
the
divergences
coming from the first terms of (25) and (26), for example.  
Since the
integration over $x$ is however not necessary as long as the
 approximation in the massless limit
is concerned, we 
omit, for a while, $\delta$ functions and set $\theta(x)=1$
in the
transformed functions.

Now, using the results, (24), (25), (26), (27), (28) and
${\cal
H}m^{-k}=x^{k}/k!$, we find 
\beg 
\hat V(\sigma, x)={1 \over 2\pi x^2}-{g\sigma  \over
\pi x}+{g^{2}\sigma^{2} \over 2\pi}(-\log\Lambda x-\gamma
+1)-\sum^{\infty}_{n=3}{(-g\sigma)^{n}x^{n-2} \over \pi
n!(n-2)}. 
\ene 
Note
that, due to the creation of $1/k!$ in ${\cal H}[1/m^{k}]$,
the series
converges
for any large $x$.  Therefore
the large $x$ behavior of $\hat V$ can be easily accessed by
increasing the order
of expansion.  This is one of the advantages of $\hat V$ over
$V$.

We turn to the approximation of the massless effective
potential by perturbative series at order $L$,
$\hat V_{L}$.  At $L$-th order, we have
just first $L+1$ terms of (29) and find
\beg
\hat V_{L}(\sigma, x)={1 \over 2\pi x^2}-{g\sigma \over
\pi x}+{g^{2}\sigma^{2} \over 2\pi}(-\log \Lambda x-\gamma
+1)-\sum^{L}_{n=3}{(-g\sigma)^{n}x^{n-2} \over \pi n!(n-2)}. 
\ene 
The input $x^{*}$ will be determined as in the previous
section.
Actually the break down appears as the domination of the last
term in
 $\hat V_{L}$ which shows up as its unlimited behavior for
large $x$.  This can
be seen in Fig. 3.  And before the domination the function
experiences a
stationary behavior for odd $L$ and large even $L$.  We find
that the plateau region represents the end of the
reliable perturbative regime and thus we fix $x^{*}$ by
the equation,
\begin{equation} 
{\partial \hat V_{L}(\sigma, x^{*}) \over \partial x^{*}}=0.
\end{equation} 
If there are several solutions we should input the largest
one into
$x^{*}$ due
to the obvious reason.

Now, the condition (31) gives $x^{*}$ as $constant/g\sigma$
for odd $L$
\renewcommand{\thefootnote}{\fnsymbol{footnote}}
{\footnote[2]{\normalsize 
For
even $L$, there is no solution for (31).  In these cases,
however, the
use
of
$\partial^{2} \hat V_{L}(\sigma, x) /\partial x^{2}=0$ gives
a solution
by which we still have good approximation  of $m_{dyn}$ at
$L=4,6,8,\cdots$.}}.
For odd $L$, the substitution of the solution into $\hat
V_{L}$ gives
the
optimized
potential $V_{opt}(\sigma)$.  For example for $L=3$, we have
the
solution,
$g\sigma x^{*}=1.59607$, and this gives the optimized
potential,
\begin{equation} 
V_{opt}={g^{2}\sigma^{2} \over 2\pi}\Bigl(\log{g\sigma \over
\Lambda}-0.373264\Bigl). 
\end{equation} 
The dynamical mass is given from
$V_{opt}$ in the standard way.  Note that, since $x$ must be
positive
(see
section 2), the region of $V_{opt}$ thus approximated is
restricted to
the
positive $\sigma$.  In the following, we summarize the result
of the
approximate
calculation of the dynamical mass for $L=3,5,7,9,11$; 
\begin{eqnarray} 
L=3,
\hskip 3mm m_{dyn}/\Lambda &=& 0.880966\hskip 3mm({\rm
at}\hskip 3pt
g\sigma
x^{*}=1.59607),\nonumber\\ 
L=5, \hskip 3mm m_{dyn}/\Lambda &=& 0.97760 \hskip
3mm({\rm at}\hskip 3pt g\sigma x^{*}=2.18061),\nonumber\\ 
L=7, \hskip 3mm
m_{dyn}/\Lambda &=& 0.99401\hskip 3mm({\rm at}\hskip 3pt
g\sigma
x^{*}=2.75900),\nonumber\\ 
L=9, \hskip 3mm m_{dyn}/\Lambda &=& 0.99809\hskip 3mm({\rm
at}\hskip 3pt g\sigma x^{*}=3.33355),\nonumber\\ 
L=11, \hskip 3mm
m_{dyn}/\Lambda &=& 0.999326 \hskip 3mm({\rm at}\hskip 3pt
g\sigma
x^{*}=3.90545). 
\end{eqnarray} 
The above result is quite good.  Thus, via Heaviside
transform approach, 
the dynamical mass is approximated only from perturbative
information.

The small mass expansion can be also obtained.  Our task is
just to
substitute the solution of (31) into the approximate
coefficients,
\beg
\alpha_{k}(x^{*})=\int^{x^{*}}_{-\infty}dx(-x)^{k}{\partial
\hat
V_{L}(\sigma,x) \over \partial  x}.
\ene
To perform integration, we need the full form of $\hat V_{L}$
including
the
$\theta$ and $\delta$ functions.  The full form is given by
\begin{eqnarray}
\hat V_{L}(x)&=&{1 \over 2\pi}\biggl[{1 \over
x^2}\theta(x)-{2 \over
x}\delta(x)-(\gamma+\log x\mu+1/2)\delta^{'}(x)\biggl]
-{g\sigma \over \pi}\biggl[{1 \over x}\theta(x)+(\gamma+\log
x\mu)\delta(x)\biggl]\nonumber\\
&-&{g^{2}\sigma^{2} \over 2\pi}(\gamma+\log
x\Lambda-1)\theta(x)-\sum^{L}_{n=3}{(-g\sigma)^n x^{n-2}
\over \pi
n!(n-2)}\theta(x).
\end{eqnarray}
From (35) $\alpha_{k}$ is given at $L$-th order as
\begin{eqnarray}
\alpha_{1}&=&{g\sigma \over \pi}\biggl[-{1 \over
X}+1-\gamma+\log(g\sigma/\mu)-\log
X+\sum_{n=2}^{L}{(-1)^nX^{n-1} \over
n!(n-1)}\biggl],\nonumber\\
\alpha_{2}&=&{1 \over \pi}\biggl[\log(g\sigma/\mu)-\log
X+2+\sum_{n=1}^{L}{(-1)^nX^{n} \over n!n}\biggl],\nonumber\\
\alpha_{k}&=&{(-1)^{k+1} \over \pi
(g\sigma)^{k-2}}\sum_{n=0}^{L}{(-1)^nX^{n+k-2} \over 
n!(n+k-2)}\quad
(k>2),
\end{eqnarray}
where $X=x^{*}g\sigma$.
At $L=11$, for example, we have
\begin{eqnarray}
V(\sigma,m)&\sim & {g^{2}\sigma^{2} \over
2\pi}\Bigl(\log{g\sigma \over
\Lambda}-0.499326\Bigl)+m{g\sigma \over \pi}(\log
g\sigma-0.00139409)+
{m^2 \over 2}{1 \over \pi}(\log g\sigma+1.00595)\nonumber\\
 &+&{m^3 \over 6g\sigma}0.973714-{m^4 \over
24g^2\sigma^2}0.878954+O(m^5).
\end{eqnarray}
This is quite accurate because the exact result reads
\beg
 V(\sigma,m)={g^{2}\sigma^{2} \over
2\pi}\Bigl(\log{g\sigma \over \Lambda}-{1 \over 2}\Bigl)+ {1
\over
\pi}mg\sigma\log{g\sigma \over
\mu}+ {m^2 \over 2\pi}(\log{g\sigma \over \mu}+1)+{m^3 \over
6g\sigma}-{m^4
\over 24g^2\sigma^2}+O(m^5).
\ene

Before closing this section, we prove that our approximants
for the
massless potential converges to the
exact result in the $L\rightarrow \infty$ limit.  That is,
$\lim_{L\rightarrow \infty}\hat V_{L}(\sigma,x^{*})=
V(\sigma,0)$. 
From
(30) we find that $\lim_{L\rightarrow \infty}\partial\hat
V_{L}/\partial
x$ can
be easily summed up and, using $\lim_{L\rightarrow \infty
}\hat
V_{L}=\hat
V$ (given by (29)),
\beg
{\partial \hat
V(\sigma, x) \over \partial x}=-{1 \over \pi x^3}\exp\Bigl[
-{g\sigma x}\Bigl].
\ene
The perturbative truncation of
(39) is given by expanding $\exp[-g\sigma x]$ to relevant
orders.  Since
$\rho=\infty$ for the series expansion of (39), the solution
of the
truncated version of $(39)=0$ approaches to $\infty$ in the
$L\rightarrow
\infty$ limit.  More precisely we find the scaling of the
solution for
large
$L$,
\beg
g\sigma x^{*} \sim {1 \over 3}L.
\ene
Now consider the reminder $\hat R_{L}$, defined by
\beg
\hat R_{L}=-\sum^{\infty}_{n=L+1}{(-g\sigma)^{n} \over \pi
n!(n-2)}x^{n-2}.
\ene
Since $\hat V_{L}+\hat R_{L}=\hat V_{\infty}$ and
$\rho=\infty$ for
$\hat
V_{\infty}$, it is sufficient to show
that
\beg
\hat R_{L}(\sigma,x^{*})\rightarrow 0\quad (L\rightarrow
\infty).
\ene
Note here that $x^{*}$ depends on $L$ and behaves at large
$L$ as (40).  
Now, using the Stirling's formula and (40), we have
\beg
|\hat R_{L}|<{1 \over x^{*}\sqrt{2\pi^3 L^{3}}}{(eg\sigma
x^*/L)^{L+1}
\over
1-eg\sigma x^*/L}<{9(g\sigma)^2 \over
\sqrt{2\pi^3}(3/e-1)}L^{-7/2}(e/3)^L\rightarrow 0\quad
(L\rightarrow
\infty),
\ene
which proves the convergence.

\vskip 20pt 
\noindent {\bf 4 Discussion} 
\vskip 10pt  

One reason of the success of our approximate calculation is
that the
transformed
function has infinite radius of convergence.  The other
reason is that,
as
the order increases, $\hat
V_{L}(\sigma,x)$ quickly approaches to the value at
$x=\infty$ for fixed
$\sigma$.  If one uses the closed form, 
\begin{equation} 
\hat V=
{g^{2}\sigma^{2} \over 2\pi}\Bigl(\log{g\sigma \over \Lambda}
-{1 \over
2}\Bigl)+{g^{2}\sigma^2 \over \pi}\int^{\infty}_{g\sigma
x}dx{e^{-x}
\over x^3}, 
\end{equation}
one finds the reason by expanding (44) for large $g\sigma x$,
\begin{equation} 
\hat V(\sigma,x)={g^{2}\sigma^{2} \over
2\pi}\Bigl(\log{g\sigma \over \Lambda} -{1
\over 2}\Bigl)+{1 \over
\pi x^{2}}\exp\biggl[-g\sigma x\biggl]\Biggl({1 \over g\sigma
x}+(-3)({1
\over
g\sigma x})^2+ (-3)(-4)({1 \over g\sigma x})^3+O(({1 \over
g\sigma x})^4)\Biggl).
\end{equation} 
By contrast, the original function has the power-like
expansion as shown
in (38).  Thus it
is obvious that the approximation of the massless potential
is more
convenient in
$\hat V_{L}$ since it approaches to the "massless" value much
faster
than $V_{L}$.  The 
reason behind why the transformed function behaves so good is
not known
to us.   

We have shown that the perturbative series at finite orders
produces the
approximate massless effective potential and the dynamical
mass.  The
deformation of the effective potential
was made
by Heaviside transform with respect to the mass and the
stationarity 
prescription to fix
the input value $x^{*}$ has found to work good.  It was also
shown that
our scheme is capable of approximating the small $m$
behavior.  Thus the Heaviside transform drastically improves
the status of the perturbative approximation of physical
quantities.  We are under
the study of full approximation method by which the general
case
where the
explicit mass is not small can be treated.  The result of
investigation will
be reported elsewhere.

\vskip 24pt {\bf Acknowledgment}

The author thanks Dr. H. Suzuki for the critics on the
primitive version
of
this
work and stimulating discussion.

\newpage \begin{center} {\bf References} \end{center}
\begin{description}
\item
[{1}] D.J.Gross and A.Neveu, Phys. Rev. D10 (1974) 3235. 
\item [{2}] S.
Moriguchi
et al, Suugaku Koushiki II, Iwanami Shoten (in Japanese). 
\item [{3}] H.
Yamada,
Mod. Phys. Lett. A11 (1996) 1001. 
\item [{4}] G. t'Hooft and M. Veltman,
Nucl.
Phys. B44 (1972) 189; \\ C. G. Bollini and J. J. Giambiagi,
Phys. Lett.
B40
(1972) 566;\\ G. M. Cicuta and E. Montaldi, Nuovo Cimento
Lett. 4 (1972)
329.
\item [{5}] W. A. Bardeen, A. J. Buras, D. W. Duke and T.
Muta, Phys.
Rev.
D18
(1978) 3998. \end{description}

\begin{center} {\bf Figure Captions} \end{center}
\begin{description} 
\item
[{Figure 1}] Model series $\hat f_{L}(x)$ is shown for
$L=1,6,11$.
\item
[{Figure 2}] The Feynman diagrams contributing to the
effective
potential at
the
leading order of $1/N$ expansion. 
\item [{Figure 3}] The Heaviside functions
of
the effective potential with fixed $\sigma$ at $L=3,6$ and
$9$.  For the
sake of
the simplicity, we have set that $g\sigma=1$ and $\Lambda=1$.
\end{description}

\end{document}